\documentclass[conference]{IEEEtran}
\IEEEoverridecommandlockouts
\usepackage{cite}
\usepackage{amsmath,amssymb,amsfonts}
\usepackage{algorithmic}
\usepackage{graphicx}
\usepackage{textcomp}
\usepackage{xcolor}
\usepackage{colortbl}
\usepackage{bm}
\usepackage{amsmath}
\usepackage{xcolor}
\usepackage{colortbl}
\usepackage{pifont}
\usepackage{booktabs} 
\usepackage{makecell} 
\usepackage{diagbox}
\usepackage{multirow} 
\usepackage{arydshln} 
\usepackage{float}
\usepackage{color}
\usepackage{stfloats}
\usepackage{paralist}
\def\BibTeX{{\rm B\kern-.05em{\sc i\kern-.025em b}\kern-.08em
    T\kern-.1667em\lower.7ex\hbox{E}\kern-.125emX}}
\usepackage[colorlinks]{hyperref}
\usepackage[pagebackref=true,breaklinks=true,colorlinks,bookmarks=false]{}

\definecolor{rowblue}{RGB}{250,235,215}   
\definecolor{rowgreen}{RGB}{209,239,191}   
\definecolor{rowgray}{RGB}{245,245,245}   
\definecolor{highgreen}{RGB}{255,0,0}
\definecolor{commentcolor}{RGB}{237,2,140}   
\definecolor{xbcolor}{RGB}{94,38,18}   
\newcommand{\xiabiao}[1]{{\color{xbcolor}#1}}

\begin{document}

\title{Universal Organizer of SAM for Unsupervised Semantic Segmentation}

\author{
\IEEEauthorblockN{Tingting Li,
Gensheng Pei, Xinhao Cai,
Qiong Wang,
Huafeng Liu
and Yazhou Yao} 
\IEEEauthorblockA{School of Computer Science and Engineering, Nanjing University of Science and Technology, Nanjing, China\\ \{litingting, peigsh, xinhao, wangq, liu.hua.feng, yazhou.yao\}@njust.edu.cn } 
}

\maketitle

\begin{abstract}
Unsupervised semantic segmentation (USS) aims to achieve high-quality segmentation without manual pixel-level annotations. Existing USS models provide coarse category classification for regions, but the results often have blurry and imprecise edges. Recently, a robust framework called the segment anything model (SAM) has been proven to deliver precise boundary object masks. Therefore, this paper proposes a universal organizer based on SAM, termed as UO-SAM, to enhance the mask quality of USS models. Specifically, using only the original image and the masks generated by the USS model, we extract visual features to obtain positional prompts for target objects. Then, we activate a local region optimizer that performs segmentation using SAM on a per-object basis. Finally, we employ a global region optimizer to incorporate global image information and refine the masks to obtain the final fine-grained masks. Compared to existing methods, our UO-SAM achieves state-of-the-art performance. Our codes are available at \textcolor{red}{https://github.com/NUST-Machine-Intelligence-Laboratory/UO-SAM}.
\end{abstract}


\section{Introduction}\label{sec:intro}
Semantic segmentation~\cite{pei2022hierarchical,pei2023hierarchical,10105896,10023953} is a critical task in computer vision that aims to assign semantic labels~\cite{tang2016tri,pei2024videomac} to each pixel in an image. It has been widely applied in various fields such as medical imaging~\cite{medical}, aerial remote sensing~\cite{Aerial}, and autonomous driving~\cite{drive}. In recent years, the success of deep learning techniques and the availability of large-scale pixel-level annotated datasets have greatly improved semantic segmentation performance. However, annotating images at the pixel level requires significant human resources, making it extremely costly. This has led to the emergence of weakly supervised~\cite{yao2021non,chen2023multi,zhang2020causal,chen2024spatial,chen2022saliency,chen2021semantically,10185466,10023953} and unsupervised \cite{chen2021enhanced,yao2020towards,sun2022pnp} semantic segmentation (USS). This paper focuses on USS, which aims to capture pixel-level semantics from unlabeled data and is one of the most challenging tasks.

\begin{figure}[t]
    \centering
    \includegraphics[width=\linewidth]{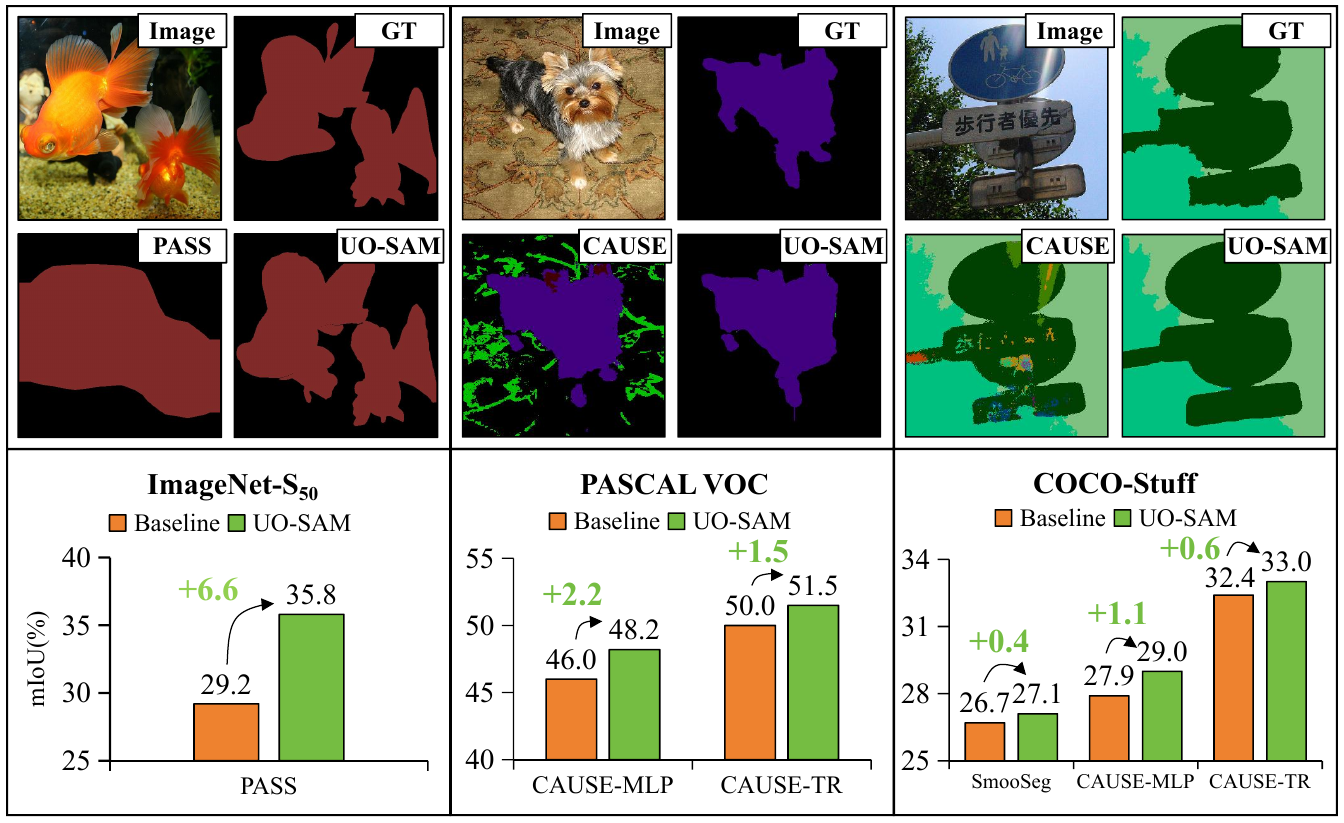}
    \caption{State-of-the-art comparison on three datasets (\textit{i.e.}, ImageNet-S$_{50}$~\cite{PASS}, PASCAL VOC~\cite{pascalvoc}, and COCO-Stuff~\cite{coco-stuff}). We present UO-SAM, which, for the first time, outperforms the top-specialized baselines on multiple USS benchmarks.}
    \label{fig:fig_first}
\end{figure}

Early USS works~\cite{zhou2023attribute,li2024learning,yao2021jo,yao2017exploiting} attempt to train models to learn semantic consistency without prior knowledge. For example, DeepCluster~\cite{DeepCluster} introduces clustering to group features, IIC~\cite{IIC} matches pairs of categories and maximizes the mutual information between them during training, and PiCIE~\cite{picie} utilizes geometric consistency as an inductive bias by clustering pixel-level features with invariance and equal variances. Given the good semantic consistency showcased by pre-training self-supervised frameworks, self-supervised ViT models (\textit{e.g.}, DINO~\cite{DINO}) have gained considerable popularity. TransFGU~\cite{transfgu} uses class activation maps~\cite{cam} as pixel-level pseudo-labels, mapping high-level semantic classes discovered in DINO to low-level pixel features. STEGO~\cite{STEGO} designs new loss functions to make same-class features more compact while preserving relationships among different categories. HP~\cite{HP} proposes hidden positive factors to maintain local semantic consistency and designs gradient propagation strategies to match them. SmooSeg~\cite{smooseg} simplifies segmentation using smoothness priors and introduces a smoothness loss to promote intra-segment smoothness while maintaining discontinuity between different segments. CAUSE~\cite{CAUSE} leverages insights from causal reasoning and proposes a causal unsupervised semantic analysis framework. LUSS~\cite{PASS} demonstrates the feasibility of large-scale USS tasks and constructs a new dataset, ImageNet-S, along with its baseline.

\begin{figure*}[t]
	\centering
	\includegraphics[width=0.91\linewidth]{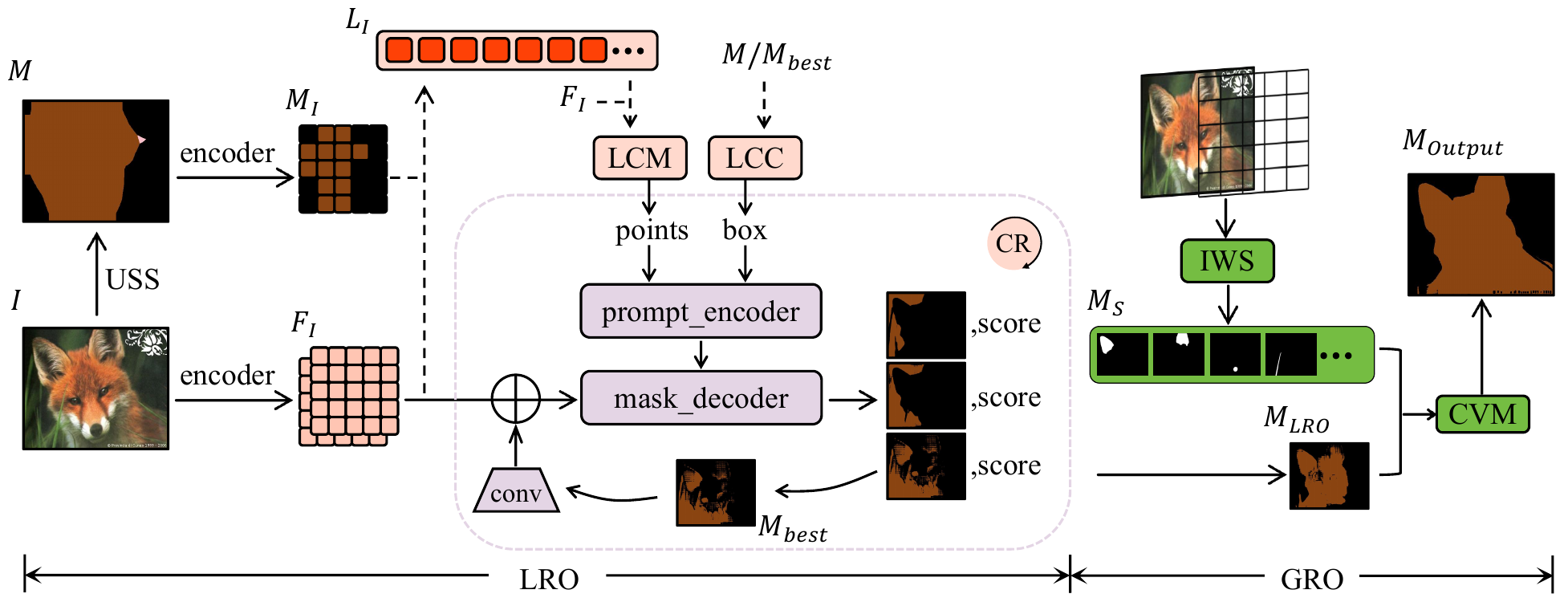}
	\caption{The architecture of our UO-SAM. We introduce a universal organizer of segment anything model (UO-SAM). It consists of the local region optimizer (LRO) and the global region optimizer (GRO) modules to improve the USS baseline performance.}
	\label{fig:fig_framework}
\end{figure*}

Unsupervised semantic segmentation presents numerous challenges~\cite{jiang2024dual}, and the quality of masks produced by existing models remains subpar due to issues such as incomplete segmentation, incorrect segmentation, and blurred boundaries. Consequently, we propose a universal organizer framework predicated on SAM~\cite{sam}, termed as UO-SAM, designed to enhance the precision and quality of masks.
Specifically, we introduce the local region optimizer (LRO) to rectify unclear mask boundaries. This module isolates specific object-related feature fragments from image features utilizing mask features, subsequently establishing a confidence map that mirrors the position of the target regions. We amalgamate the reliable position prompts with SAM optimization to generate masks with precise boundaries.
Additionally, to ensure the completeness of target objects in multi-label scenarios, we propose the global region optimizer (GRO). This module harnesses contextual information from the image, focuses on all target objects, and engineers a semantic voting mechanism to better integrate the predictions from the LRO module. This results in a reduction of erroneous pixels in the masks and yields enhanced segmentation results.

As shown in Fig.~\ref{fig:fig_first}, we present extensive experimental evaluations conducted on multiple USS datasets. Building upon established unsupervised semantic segmentation models like PASS~\cite{PASS}, CAUSE~\cite{CAUSE}, and SmooSeg~\cite{smooseg}, UO-SAM delivers remarkable achievements across diverse settings. 
Notably, our UO-SAM surpasses existing state-of-the-art approaches. Specifically, on ImageNet-S$_{50}$~\cite{PASS}, our approach exhibits superior performance compared to PASS with a 6.6\% improvement. UO-SAM improves CAUSE-MLP and CAUSE-TR by 2.2\% and 1.5\% on PASCAL VOC~\cite{pascalvoc}. Our approach outperforms SmooSeg, CAUSE-MLP and CAUSE-TR on COCO-Stuff~\cite{coco-stuff} with performance improvements of 0.4\%, 1.1\%, and 0.6\%, respectively. Our contributions can be summarized as follows:

(1) We propose a universal organizer framework based on SAM (UO-SAM), which consists of two modules: the local region optimizer (LRO, see \S\ref{sec:LRO}) and the global region optimizer (GRO, see \S\ref{sec:GRO}).

(2) We introduce LRO to alleviate the issues of blurry and unclear mask boundaries along object edges. Moreover, we present GRO to ensure the accuracy and completeness of overall segmentation, thereby minimizing incomplete segmentation instances.

(3) Our UO-SAM improves the segmentation accuracy of multiple baseline models by a sizable amount and requires no additional re-training or fine-tuning.

\section{Proposed Method}

In this section, we will introduce our UO-SAM. We begin by providing an overview of the framework in \S\ref{sec:overall}. Then, in \S\ref{sec:LRO}, we present the local region optimizer module for optimizing object segmentation by leveraging foreground position prompts. Additionally, we discuss the global region optimizer module in \S\ref{sec:GRO}.

\subsection{Overall Framework}\label{sec:overall}

In this work, we present the universal organizer of segment anything model (UO-SAM) for the USS task. As illustrated in Fig.~\ref{fig:fig_framework}, our approach integrates two primary components: the local region optimizer (LRO) and the global region optimizer (GRO). Following the structure of SAM~\cite{sam}, our UO-SAM is constructed around three key elements: an image encoder ($\texttt{Enc}_I$), a prompt encoder ($\texttt{Enc}_P$), and a lightweight mask decoder ($\texttt{Dec}_M$).

Upon processing an image through the image encoder, an image embedding is generated which can be effectively interrogated by various input prompts to yield object masks. Within the LRO component, the location confidence map (LCM) and the largest connected component (LCC) are employed to identify point and box prompts. Consequently, optimized masks for target objects are acquired through cascaded post-processing. Subsequently, the GRO module utilizes a semantic voting mechanism to execute full image optimization on it, thereby generating the final mask.

\subsection{Local Region Optimizer}\label{sec:LRO}
The user only needs to provide an image $I$ and the mask $M$ generated by an unsupervised semantic segmentation model. UO-SAM autonomously localizes the positional information of the target object and feeds it into the prompt encoder in the form of points, bounding boxes, or other visual prompts. Given the image $I$ and the mask $M$, we apply the image encoder of SAM to extract their visual features, represented as:
\begin{equation}
F_I=\texttt{Enc}_I(F), M_I=\texttt{Enc}_I(M),
\end{equation}
where $F_I \in \mathbb{R}^{H \times W\times C}$, $M_I \in \mathbb{R}^{H \times W\times 1}$, with $H$, $W$ denoting the spatial dimension of the image feature map and $C$ indicating the channels. Then, by utilizing $M_I$ to crop the foreground pixel features of the target object from $F_I$, we obtain a set of local features expressed as:
\begin{equation}
\left\{ L^i_I \right\}^n_{i=1}= M_I \circ F_I,
\end{equation}
where $\left\{ L^i_I \right\}^n_{i=1} \in \mathbb{R}^{1 \times C}$, $n$ is the number of non-zero mask elements in $M_I$, and ${\circ}$ indicates the space multiplication.

\noindent\textbf{Location Confidence Map (LCM).} For extracting foreground position prompts (points/boxes), we first compute the similarity ($\texttt{sim}$) between the image features ($F_I$) and the foreground features ($\left\{ L^i_I \right\}^n_{i=1}$) to obtain $n$ confidence maps, denoted as $s^i$. These confidence maps provide information about the likelihood of each pixel being part of the foreground. Next, we aggregate these confidence maps using average pooling to obtain $S$, which integrates the distribution probabilities of different parts of the target object,
\begin{equation}
s^i = \texttt{sim}(F_I, L^i_I), S=\dfrac{1}{n} {\sum^n_{i=1}} s^i,
\end{equation}
where $s^i,S \in \mathbb{R}^{H \times W}$. Based on the overall position estimation, we select the highest-scoring point ($P_{pos}$) and the lowest-scoring point ($P_{neg}$) in $S$ as foreground and background points, respectively. The prompt encoder then utilizes these points to guide SAM in focusing its attention on the central position of the target object for segmentation while disregarding potential background regions, represented as:
\begin{equation}
T_P = \texttt{Enc}_P (P_{pos},P_{neg}).
\end{equation}

\noindent\textbf{Largest Connected Component (LCC).} To ensure the integrity of the segmented object, we search for the largest connected component $\mathbb{L}_{cc}$ on the mask $M$ that contains $P_{pos}$. We can expand the captured foreground region of the target object by identifying $\mathbb{L}_{cc}$, 
\begin{equation}
\mathbb{L}_{cc}=\texttt{max} \left\{ l|P_{pos} \in l \right\}.
\end{equation}
During the segmentation process, the prompt encoder $Enc_P$ incorporates bounding boxes as visual cues to help direct the model's attention towards more complete object center regions, it can be described as:
\begin{equation}
T_B = \texttt{Enc}_P(B).
\end{equation}

\noindent\textbf{Cascaded Refinement (CR).} After the aforementioned steps, we obtain three initial segmentation masks of the target object and their corresponding confidence scores from the decoder of SAM. Despite selecting the mask with the highest score, it may still contain coarse edges and isolated background noise. To further refine the segmentation, we iteratively embed dense cues (\textit{i.e.}, masks) using convolutions and add them to the image embeddings.

Therefore, we design two post-processing operations. Firstly, we prompt the decoder using positive-negative point cues and the largest connected component of the original mask. Secondly, we use the coarse mask obtained from the first step and its bounding box ($T^{'}_B$) to prompt the decoder for more accurate target localization, formalized as:
\begin{equation}M_{firet\_step} = \texttt{Dec}_M(F_I,\texttt{Concat}(T_P,T_B)),
\end{equation}
\begin{equation}T_M = F_I \oplus \texttt{Conv}(M_{firet\_step}),
\end{equation}
\begin{equation}M_{second\_step} = \texttt{Dec}_M(T_M,\texttt{Concat}(T_P,T^{'}_B )),
\end{equation}
where $\texttt{Concat}$ denotes concatenation operation, and $\oplus$ indicates matrix addition.

\subsection{Global Region Optimizer}\label{sec:GRO}
In \S\ref{sec:LRO}, LRO potentially causes the model to focus on the target region and ignore other potential objects or background information. This also limits the model's ability to consider the contextual information of the image, thereby restricting the overall accuracy and completeness of the segmentation results. Therefore, we introduce GRO to segment the entire image range, not just focusing on specific target objects.

\noindent\textbf{Image-Wide Segmentation (IWS).} We first define the number of sampling points $N$ for each side of the input image. Based on this, we generate a two-dimensional grid of points with a uniform and evenly spaced distribution on the plane. Thus, the point set $\mathbb{P}$ is denoted as:
\begin{equation}
\mathbb{P}=\left \{ (x,y)|x=o+(i*g),
y=o+(j*g)
\right\},
\end{equation}
where $o$ represents the offset, $g$ represents the spacing between points, $i,j\in N$. Next, we generate a series of crop boxes of different sizes for the original image. The grid points on each crop box represent the regions of interest, guiding the segmentation process. We use the non-maximum suppression (NMS) for the generated segmentation fragments to perform quality filtering and remove duplicates. Finally, the processed mask data is concatenated together to generate high-quality masks $M_S^i$ for all objects in the image. This approach helps improve the overall accuracy and quality of the segmentation.

\noindent\textbf{Category Voting Mechanism (CVM).} To integrate a set of high-quality masks $M_S^i$ generated by image-wide segmentation, which lack category information with the locally optimized mask $M_{LRO}$ obtained from the LRO module, we design a category voting mechanism. For each $M_S^i$, we overlay it with the mask $M_{LRO}$ and extract the corresponding pixel categories. We then calculate the class ID with the highest vote for each pixel position. This top-voted class ID is selected as the proposal and applied to the pixels corresponding to the valid mask. By employing this mechanism, we can obtain a fused mask that has boundaries as accurately as possible and contains minimal erroneous pixels.

\section{Experiment}

\subsection{Experiment Setup}

\noindent\textbf{Datasets.} To validate the efficacy of UO-SAM, we utilize three USS datasets, including ImageNet-S$_{50}$~\cite{PASS}, PASCAL VOC~\cite{pascalvoc}, and COCO-Stuff ~\cite{coco-stuff}. The ImageNet-S$_{50}$ dataset, derived from the ImageNet-1k~\cite{deng2009imagenet} dataset, encompasses 64,431 training images, 752 validation images, and 1,682 test images across 50 categories, excluding the background category. Precise pixel-level annotations are provided for both the validation and test sets. PASCAL VOC, a prevalent dataset in semantic segmentation, comprises 21 object categories, including 20 foreground object categories and one background category. Lastly, the COCO-Stuff dataset, a scene texture segmentation subset of MS-COCO 2017~\cite{lin2014microsoft}, provides comprehensive pixel-level annotations for objects/things categories.

\noindent\textbf{Statistical Analysis.} Table~\ref{tab:categories} delineates the category counts per image for the validation sets of ImageNet-S$_{50}$~\cite{PASS}, PASCAL VOC~\cite{pascalvoc} and COCO-Stuff~\cite{coco-stuff}. Within the ImageNet-S$_{50}$ dataset, the majority of the images contain one category, while 0.93\% of the images contain more than one category. Regarding PASCAL VOC, approximately 37\% of the images are single-category, meaning they only consist of one category. Furthermore, around 50\% of the images have two categories, indicating the presence of two target objects or items. Contrarily, most images within COCO-Stuff contain five categories. These statistics suggest from ImageNet-S$_{50}$ to PASCAL VOC, and onward to COCO-Stuff, the complexity of categories gradually increases across these three datasets.

\noindent\textbf{Evaluation Metrics.} For ImageNet-S$_{50}$~\cite{PASS}, we adopt the evaluation standards of PASS~\cite{PASS}, using mean intersection over union (mIoU), boundary mIoU (b-mIoU), image-level accuracy (Img-Acc), and F-measure(${F_\beta} $) as metrics. During the evaluation process, all images are evaluated at their original resolutions. mIoU and b-mIoU serve as comprehensive evaluation metrics, while Img-Acc and $F_\beta$ provide insights into performance in terms of both category and shape. For PASCAL VOC~\cite{pascalvoc} and COCO-Stuff~\cite{coco-stuff}, we follow the protocol of CAUSE~\cite{CAUSE} and utilize two common metrics: pixel accuracy (pAcc) and mIoU to measure performance.

\noindent\textbf{Implementation Details.} Our proposed method is implemented based on the PyTorch framework~\cite{paszke2019pytorch}, and all experiments are executed on an NVIDIA GeForce RTX 3090 GPU. To ensure a fair comparison with baselines~\cite{PASS,CAUSE,smooseg}, we adopt their weight inference-based label generation approach. Subsequently, we optimize the generated labels employing our proposed UO-SAM method and evaluate the results using the evaluation metrics outlined in their respective works. It is worth noting that UO-SAM does not require a training process and can be used to optimize the generated label files independently or embedded within the evaluation code. The inference time per image is roughly 1.26 seconds.

\begin{table}[t]
    \renewcommand\arraystretch{1.2}
    \centering
    \caption{The number of categories per image in the validation sets of ImageNet-S$_{50}$, PASCAL VOC, and COCO-Stuff.}
    \label{tab:categories}
    \setlength{\tabcolsep}{3.5pt}
    \resizebox{\columnwidth}{!}{
    \begin{tabular}{lcccccc}
    \hline
    & \multicolumn{6}{c}{\textbf{Number of images on validation set}}\\       
    \hline
    \textbf{Categories in each image} &1 &2 &3 &4 &5 & $>5$ \\					 
    \hline
    ImageNet-S$_{50}$~\cite{PASS} &745 &7 &0 &0 &0 &0  \\
    PASCAL VOC~\cite{pascalvoc} &544 &729 &176 &0 &0 &0  \\
    COCO-Stuff~\cite{coco-stuff} &0 &0 &6 &199 &1,970 &0 \\
    \hline
    \end{tabular}}
\end{table}

\begin{table}[t]
    \renewcommand\arraystretch{1.2}
    \centering
    \caption{Comparisons with existing USS approaches on the ImageNet-S$_{50}$ validation set. `Sal' denotes the method employs saliency maps, and `IN1k' denotes initialization with supervised pre-training on ImageNet-1k.}
    \label{tab:ImageNet}
    \resizebox{\linewidth}{!}{
    \setlength\tabcolsep{2.5pt}
    \begin{tabular}{lccccc}
    \hline
    \textbf{Method} &\textbf{Pior} &\textbf{mIoU} &\textbf{B-mIoU} &\textbf{Img-Acc} &\bm{${F_\beta}$}   \\
    \hline
    DeepCluster\xiabiao{${\rm_{ECCV18}}$}~\cite{DeepCluster} &- &4.0 &1.4 &14.9 &31.6 \\
    DeepCluster~\cite{DeepCluster} &IN1k &14.6  &3.1  &44.8 &33.2  \\
    PiCIE\xiabiao{${\rm_{CVPR21}}$}~\cite{picie} &- &5.0 &1.8 &15.8 &14.6  \\
    PiCIE~\cite{picie} &IN1k &17.8 &3.7 &45.0 &32.1  \\
    MaskCon\xiabiao{${\rm_{ICCV21}}$}~\cite{maskcontrast} &Sal &24.6 &15.6 &47.9 &\textbf{65.7}  \\
    \hdashline
    PASS\xiabiao{${\rm_{TPAMI23}}$}~\cite{PASS} &- &29.2 &7.6 &\textbf{66.2} &49.0  \\
    \rowcolor{rowblue} \textbf{UO-SAM (PASS)} &-&\textbf{35.8}\textcolor{highgreen}{$_{\uparrow{6.6}}$} &\textbf{25.7} &64.6 &56.7 \\
    \hline
    \end{tabular}}
\end{table}

\begin{table}[t]
    \renewcommand\arraystretch{1.2}
    \centering
    \caption{Comparisons with existing USS approaches on the PASCAL VOC validation set.}
    \label{tab:pascalvoc}
    \setlength{\tabcolsep}{8.5pt}
    \resizebox{\linewidth}{!}{
    \setlength\tabcolsep{6.5pt}
    \begin{tabular}{lccccccccc}
    \hline
    \textbf{Method} &\textbf{Backbone} &\textbf{mIoU} &\textbf{pAcc} \\
    \hline
    MaskCon\xiabiao{${\rm_{ICCV21}}$}~\cite{maskcontrast}&ResNet50&35.0&- \\
    TransFGU\xiabiao{${\rm_{ECCV22}}$}~\cite{transfgu}&ViT-S/8&37.2&-\\
    ACSeg\xiabiao{${\rm_{CVPR23}}$}~\cite{ACSeg}&ViT-S/8&47.1&52.7\\
    \hdashline
    CAUSE-MLP\xiabiao{${\rm_{arxiv23}}$}~\cite{CAUSE}&ViT-S/8&46.0&85.1\\
    \rowcolor{rowblue}\textbf{UO-SAM (CAUSE-MLP)} &ViT-S/8 &48.2\textcolor{highgreen}{$_{\uparrow{2.2}}$} &86.0 \\
    CAUSE-TR\xiabiao{${\rm_{arxiv23}}$}~\cite{CAUSE}&ViT-S/8 &50.0 &87.6\\
    \rowcolor{rowblue}\textbf{UO-SAM (CAUSE-TR)} &ViT-S/8 &\textbf{51.5}\textcolor{highgreen}{$_{\uparrow{1.5}}$} &\textbf{88.1} \\
    \hline
    \end{tabular}}
\end{table}

\begin{table}[t]
    \renewcommand\arraystretch{1.2}
    \centering
    \caption{Comparisons with existing USS approaches on the COCO-Stuff validation set.}
    \label{tab:coco}
    \setlength{\tabcolsep}{8.0pt}
    \resizebox{\columnwidth}{!}{
    \setlength\tabcolsep{5.0pt}
    \begin{tabular}{lccccccccc}
    \hline
    \textbf{Method} &\textbf{Backbone} &\textbf{mIoU} &\textbf{pAcc} \\
    \hline
    DINO\xiabiao{${\rm_{ICCV21}}$}~\cite{DINO}&ViT-S/8&11.3&28.7 \\
    ACSeg\xiabiao{${\rm_{CVPR23}}$}~\cite{ACSeg}&ViT-S/8&16.4&-\\
    TransFGU\xiabiao{${\rm_{ECCV22}}$}~\cite{transfgu}&ViT-S/8&17.5&52.7 \\
    STEGO\xiabiao{${\rm_{ICLR22}}$}~\cite{STEGO}&ViT-S/8&24.5&48.3 \\
    HP\xiabiao{${\rm_{CVPR23}}$}~\cite{HP}&ViT-S/8&24.6&57.2 \\
    SmooSeg\xiabiao{${\rm_{NeurIPS23}}$}~\cite{smooseg}&ViT-S/8&26.7&63.2 \\
    \hdashline
    CAUSE-MLP\xiabiao{${\rm_{arxiv23}}$}~\cite{CAUSE}&ViT-S/8&27.9&66.8 \\
    \rowcolor{rowblue}\textbf{UO-SAM (CAUSE-MLP)} &ViT-S/8 &29.0\textcolor{highgreen}{$_{\uparrow{1.1}}$} &67.8\\
    CAUSE-TR\xiabiao{${\rm_{arxiv23}}$}~\cite{CAUSE}&ViT-S/8 &32.4 &69.6 \\
    \rowcolor{rowblue}\textbf{UO-SAM (CAUSE-TR)} &ViT-S/8 &\textbf{33.0}\textcolor{highgreen}{$_{\uparrow{0.6}}$} &\textbf{70.1}\\
    \hline
    \end{tabular}}
\end{table}

\begin{figure*}[t]
    \centering
    \includegraphics[width=1\linewidth]{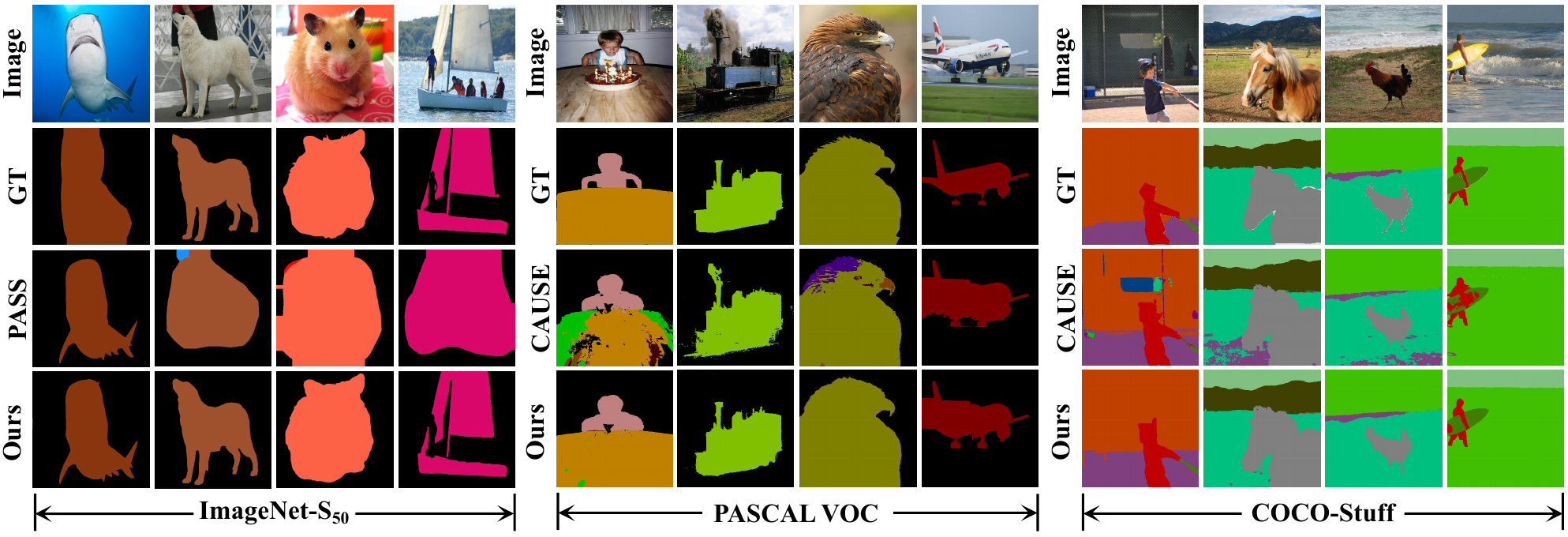}
    \vspace{-0.6cm}
    \caption{Qualitative comparisons of USS on the validation sets of ImageNet-S$_{50}$~\cite{PASS}, PASCAL VOC~\cite{pascalvoc} and COCO-Stuff~\cite{coco-stuff}.}
    \vspace{-0.3cm}
    \label{fig:fig_Imagenet50}
\end{figure*}

\subsection{Comparison to State-of-the-Arts}
In this section, we conduct comprehensive experiments on the ImageNet-S$_{50}$, PASCAL VOC, and COCO-Stuff datasets to compare our UO-SAM with state-of-the-art USS approaches. 

\noindent\textbf{Results on ImageNet-S$_{50}$.} Table~\ref{tab:ImageNet} reports the results of existing state-of-the-art USS methods on the validation set of ImageNet-S$_{50}$. From the table, it can be observed that both mIoU and B-mIoU achieve the best results on both the validation and test sets. While there is a slight decrease in Img-Acc compared to PASS, the ${F_\beta}$ score is lower than MaskCon~\cite{maskcontrast}  but significantly higher than PASS~\cite{PASS}. This indicates that UO-SAM achieves a significant improvement in mask quality at the expense of a minor loss in Img-Acc, resulting in excellent performance in the overall mIoU metric.

\noindent\textbf{Results on PASCAL VOC.} To verify the generalization of our approach, we further conduct experiments on the PASCAL VOC validation set, which has a slightly higher category complexity. As shown in Table~\ref{tab:pascalvoc}, our UO-SAM achieves the state-of-the-art results for unsupervised semantic segmentation.For example,compared to the CAUSE-MLP and CAUSE-TR methods, we achieved improvements of 2.2\% and 1.5\% on mIoU respectively.

\noindent\textbf{Results on COCO-Stuff.} Additionally, we also compare the experimental performance on COCO-Stuff, and the results are shown in Table~\ref{tab:coco}. Our UO-SAM achieves a 1.1\% improvement compared to CAUSE-MLP~\cite{CAUSE}, confirming the effectiveness of our approach on a dataset with higher category complexity. Similarly, our superior results validate the effectiveness of our UO-SAM in optimizing mask quality.

\noindent\textbf{Qualitative Results.} Qualitative results of our UO-SAM on three USS datasets are shown in Fig.~\ref{fig:fig_Imagenet50}. For ImageNet-S$_{50}$, our method exhibits significantly clearer mask boundary contours compared to PASS. Furthermore, on both the PASCAL VOC and COCO-Stuff datasets, our approach yields fewer erroneous predicted pixels compared to CAUSE~\cite{CAUSE}, while maintaining excellent boundaries. These findings further demonstrate the superiority of our method.

\subsection{Ablation Studies}
In this section, the effectiveness of the LRO (\S\ref{sec:LRO}) and GRO (\S\ref{sec:GRO}) modules is assessed through comprehensive ablation studies. The quantitative results of our UO-SAM on three USS benchmarks are enumerated in Table ~\ref{tab:ablation}. 

\noindent\textbf{LRO.} It is noted that applying the LRO module for specific region optimization of masks consistently improves the baseline performance across all three datasets. There is a performance improvement of 3.9\% for ImageNet-S$_{50}$, 0.9\% for PASCAL VOC, and 0.5\% for COCO-Stuff. Upon comparing the degrees of improvement, we can conclude that the LRO module had the most significant impact on optimizing single-category masks, and its gain gradually diminished as the number of categories per image increased. 

\begin{table}[t]
	\renewcommand\arraystretch{1.2}
	\centering
	\caption{Ablation results of our components (LRO and GRO) on ImageNet-S$_{50}$, PASCAL VOC, and COCO-Stuff.}
	\label{tab:ablation}
	\setlength{\tabcolsep}{3.5pt}
	\resizebox{\linewidth}{!}{
		\begin{tabular}{c|c|ccc|c}
			\hline
			\textbf{Dataset} &\textbf{Method} &\textbf{Baseline} &\textbf{LRO} &\textbf{GRO} &\textbf{mIoU}\\        
			\hline
			\multirow{3}{*}{ImageNet-S$_{50}$}&\multirow{3}{*}{PASS} &\ding{51} & & &29.6\\
			& &\ding{51} &\ding{51} & &33.5\\
			& &\ding{51} &\ding{51} &\ding{51} &\textbf{35.8}\\
			\hline
			\multirow{3}{*}{PASCAL VOC}&\multirow{6}{*}{CAUSE-
				MLP}&\ding{51} & & &46.0\\
			& &\ding{51} &\ding{51} & &46.9\\
			& &\ding{51} &\ding{51} &\ding{51} &\textbf{48.2}\\
			\cline{1-1}\cline{3-6}
			\multirow{3}{*}{COCO-Stuff}& &\ding{51} & &  &27.9\\
			& &\ding{51} &\ding{51} & &28.4\\
			& &\ding{51} &\ding{51} &\ding{51} &\textbf{29.0}\\
			\hline
		\end{tabular}%
	}
\end{table}

\noindent\textbf{GRO.} Additionally, after incorporating our GRO module, we observe a consistent performance improvement of 2.3\% on ImageNet-S$_{50}$, 1.3\% on PASCAL VOC, and 0.6\% on COCO-Stuff. This indicates that the GRO module improves the results of multi-category masks.

\noindent\textbf{Discussion.} We conclude that the LRO module is more concerned with object segmentation and weak in capturing global information. In contrast, the GRO module takes a holistic approach by considering various labeled objects but lacks fine-grained segmentation for each object. Therefore, integrating these two modules into the UO-SAM can improve segmentation performance on datasets with simpler categories and impact datasets with more complex categories.

\section{Conclusion}
In this work, we present a SAM-based universal framework, termed as UO-SAM, to enhance the mask quality of USS models. Specifically, we propose a local region optimizer module for fine-tuning mask-specific objects. This module utilizes extracted visual features to generate a confidence map for object localization, obtains positive and negative point cues, and identifies boundary cues by finding the largest connected component, enabling cascade optimization. Further, considering the challenge of varying image categories, we introduce a global region optimizer module intended to escalate mask reliability. A voting mechanism is designed to integrate the predicted masks from both branches, thereby generating the final segmentation map. Experimental results demonstrate the generality and effectiveness of UO-SAM, outperforming existing state-of-the-art methods on multiple USS datasets.

\bibliographystyle{IEEEtran}
\bibliography{ref}

\end{document}